\documentclass[letter,twocolumn]{jpsj3}
\usepackage{graphicx,amsmath,amssymb,bm}

\allowdisplaybreaks 

\usepackage{color}
\definecolor{Green}{rgb}{0,0.7,0}
\newcommand{\e}{ {\rm e}}

\newcommand{\suzum}[1]{\textcolor{black}{#1}}
\newcommand{\suzu}[1]{\textcolor{black}{#1}}
\newcommand{\fred}[1]{\textcolor{black}{#1}}
\def \bmk{{\bm{k}}}
\def \bmq{{\bm{q}}}
\begin{document}

\title{
Inversion Symmetry and  Wave-Function-Nodal-Lines of Dirac Electrons in Organic Conductor $\alpha$-(BEDT-TTF)$_2$I$_3$ 
}
\author{
Fr\'ed\'eric \textsc{Pi\'echon}$^{1}$ 
and
Yoshikazu \textsc{Suzumura}$^{2}$ 
}
\inst{
$^1$Laboratoire de Physique des Solides, CNRS UMR 8502, Universit\'e Paris-Sud, F-91405 Orsay Cedex, France \\
$^2$
Department of Physics, Nagoya University, 
 Chikusa-ku, Nagoya 464-8602, Japan 
}

\abst{By examining  organic conductor $\alpha$-(BEDT-TTF)$_2$I$_3$ 
 which is described 
by a nearest neighbors tight-binding model
it is shown that because of inversion symmetry,
each component of a wave function (WF) exhibits nodal lines (NLs) in the Brillouin zone.
In the absence of any band crossing, each NL connects two time reversal invariant momenta (TRIM) as partners.
In the presence of a pair of Dirac points (band crossing), 
for each band that crosses and for each WF component there is
a NL that connects the pair of Dirac points via a TRIM without
 partner.
This second kind of NL leads to a discontinuous sign change for non vanishing components of the WF. 
Such a property is at the origin of the $\pm \pi$ Berry phase accumulated 
on a contour integral encircling one Dirac point.
The results are examplified by numerical calculation of WFs components for the above conductor with  a 3/4 filled band.
}

\kword{Berry phase, nodal line, Dirac point, $\alpha$-(BEDT-TTF)$_2$I$_3$, 
inversion symmetry}


\maketitle

\begin{figure}
  \centering
\includegraphics[width=7cm]{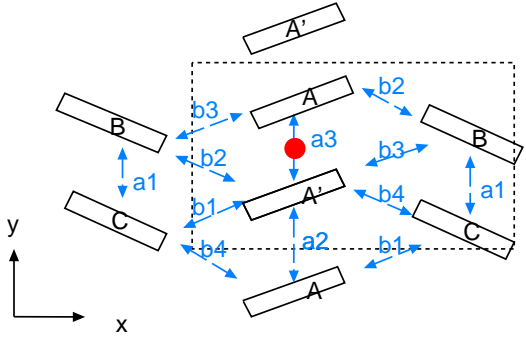}   
  \caption{(Color online)
Structure of $\alpha$-(BEDT-TTF)$_2$I$_3$ conducting plane,
 with four molecules   A, A', B, and C per unit cell (dotted square). 
The seven nearest-neighbors transfer energies  $ a_1 , \cdots, b_4$ 
 are indicated. Possible inversion centers are site B, site C as well as
the middle points between A and A' sites.\cite{Mori1984_CL}.
}
\label{fig:structure}
\end{figure}

Among  organic conductors consisting of conducting planes separated by anyon layers, 
 $\alpha$-(BEDT-TTF)$_2$I$_3$ presents a structure 
 in which the unit cell contains 
 four inequivalent molecules A, A',B, and C (see Fig \ref{fig:structure}) 
  with an inversion symmetry between A and A'.
 \cite{Mori1984_CL}
 The importance of this inversion symmetry is a central topic
 since the discovery of Dirac electrons \cite{Katayama2006_JPSJ75}. 
%
This is not surprising since this idea of Dirac points, 
 that give rise to a zero gap state at the Fermi level under pressure,
has enabled much progress in the understanding of the physical properties of $\alpha$-(BEDT-TTF)$_2$I$_3$
\cite{Tajima2009_STAM10,Kobayashi2009_STAM10}. 

Very recently it has been shown how inversion symmetry and Fu-Kane \cite{Fu2007_PRB76} topological argument 
allow to establish explicit conditions for the existence of Dirac points 
based on the sole knowledge of energy and  inversion parity eigenvalues at the four time reversal invariant momenta (TRIM) \cite{Piechon2013_JPSJ,Mori2013_JPSJ}. 
Even though such topological argument is useful to assess for the existence of Dirac touching points between valence and conduction bands, 
 it does not provide any information on their location in the Brillouin zone (BZ) .
In fact, in contrast to graphene \cite{Ando2005_JPSJ74}, the location of 
Dirac  points in $\alpha$-(BEDT-TTF)$_2$I$_3$ 
depends on pressure through the tight-binding parameters values:
in other words, Dirac  points appear at accidentally degenerated momenta in the BZ.
As a result, except for a simplified case \cite{Mori2010_JPSJ}, until now the explicit location of Dirac points is achieved 
by exploration of the full BZ either numerically
through the computation of eigen energy bands and Berry curvature\cite{Suzumura2011_JPSJ_Berry}
or using a newly developped semi-analytical method \cite{Suzumura2013_JPSJ}.

In a recent numerical study of wave functions (WFs) properties for the conduction band, it was however remarked that 
in the presence of Dirac points, the spectral functions on B and C molecules exhibit nodal lines (NLs)
(i.e. lines in the BZ where the B and C components of the WF  vanish) that connect the pair of Dirac points $\pm {\bf k}_0$
 via a TRIM \cite{Katayama2009_EPJB57,Kobayashi2013_JPSJ}. 
It was further argued that such NLs might explain the local magnetic properties on B and C molecular sites 
that are measured by NMR\cite{Takano2010_JPSJ79,Hirata2012}.

The object of this Letter is twofold. 
First, by going in an appropriate Bloch basis, we explain how the existence of NLs is intimately related to the inversion symmetry.
Second we explore the hidden properties of these NLs. We find two classes of NLs. A first class is NLs that connect two partners TRIM.
The other class emerges when there is a Dirac pair $\pm {\bf k}_0$ of touching points between two bands. For each WF component of these bands, 
there appears a NL that connects the Dirac points $\pm {\bf k}_0$ via a TRIM without partner.
It is demonstrated that one of these second type of NLs is also the location 
of a discontinuous sign change for the non vanishing components of the WF.  
This property is further related to the $\pm \pi$ Berry phase that is accumulated 
in performing a contour integral encircling one Dirac point.



We describe electronic properties of each $\alpha$-(BEDT-TTF)$_2$I$_3$ conducting plane by a tight-binding Hamiltonian 
with seven distinct nearest neighbors transfer intergrals between the four molecules A,A',B,C of the unit cell (see Fig.~\ref{fig:structure}). 
This model preserves inversion symmetry; possible inversion centers are sites B, sites C or the middle points 
of bonds AA'.
We consider a molecular Bloch basis ($|A \bmk \rangle,|A' \bmk \rangle, |B \bmk \rangle,|C \bmk \rangle $) such that the $4 \times 4$ Bloch Hamiltonian matrix reads
\begin{equation}
H_0(\bm{k})= 
\begin{pmatrix}
0 & a&b&c\\
a^*&0 &b^* e^{i k_x} & c^* e^{i k_x+i k_y}\\
b^* & b e^{-i k_x} & 0 &d\\
c^* & c e^{-i k_x-ik_y} &{d}^*&0
\end{pmatrix} \ , 
\label{eq:simple_H}
\end{equation}
with $a=a_3 + a_2 {\rm e}^{i k_y}$, $b = b_3 + b_2 {\rm e}^{i k_x}$,   
$c={\rm e}^{ i k_y}(b_4+b_1 {\rm e}^{i k_x})$ and $d=a_1 (1+ {\rm e}^{i k_y})$ (the lattice constant is taken as unity).
This matrix verifies 
time reversal symmetry $H_0(-{\bmk})=H_0^*({\bmk})$ and Bloch periodicity 
$H_0({\bmk}+\bm{G})=H_0({\bmk})$ with ${\bm{G}}$ 
a vector of the reciprocal lattice. 

To start with, we rewrite eq.~(\ref{eq:simple_H}) as $H_P(\bmk)$ in the symmetric Bloch basis $|j \bmk\rangle$ ($j=1,2,3,4$) defined by 
\begin{equation}
\begin{array}{l}
 |1 \bmk \rangle = \frac{|A \bmk \rangle+|A' \bmk \rangle}{\sqrt{2}} \ , \ \
 |2 \bmk \rangle = \frac{|A \bmk \rangle-|A' \bmk \rangle}{\sqrt{2}} \ , \\
|3 \bmk \rangle =|B \bmk \rangle \ , \ \ |4 \bmk \rangle =|C \bmk \rangle.
\end{array}
\label{eq:another_base}
\end{equation}

In this basis, inversion symmetry $\hat P$ is described by the $4 \times 4$ diagonal matrix $P(\bmk)$, with diagonal elements 
 $p_{j}(\bm{k}) = 
1, -1,   \e^{- \rm i k_x}$, and $\e^{- \rm i k_x - \rm i k_y}$, 
 for $j$ = 1, 2, 3, and 4;
such that $P(\bmk)^{-1}H_P(\bmk)P(\bmk)=H_P(-\bmk)$ and $[P(\bm{G}/2),H_P(\bm{G}/2)]=0$ at the four TRIM 
given by ${\bf \Gamma}=(0,0), {\bf X}=(\pm \pi,0), {\bf Y}=(0,\pm \pi)$ and ${\bf M}=(\pm \pi,\pm \pi)$.
For our purpose, it is convenient to perform a further unitary transformation such as to obtain a real symmetric Bloch Hamiltonian matrix:
\begin{eqnarray}
  {H}(\bm{k})  
 = {P}(\bm{k})^{-1/2}   {H}_S(\bm{k})  {P}(\bm{k})^{1/2},
 \; 
\label{real_Hamiltonian}
\end{eqnarray} 
that still obeys $[P(\bm{G}/2),H(\bm{G}/2)]=0$. 
The matrix $ {H}(\bm{k})$ appears to have only eight distinct nonvanishing elements $h_{ij}(\bmk) = (H (\bmk))_{ij}$ given by:   
$h_{11} = a_3+a_2 \cos k_y$, $h_{22} = -h_{11}$ ,
$h_{12} =  -  a_2 \sin k_y$, 
$h_{13} = \sqrt{2}(b_2 + b_3) \cos (k_x/2)$,
$h_{14} = 
           \sqrt{2} b_1 \cos (\frac{k_x+k_y}{2})
          +  \sqrt{2} b_4 \cos (\frac{k_x-k_y}{2})$,
$h_{23} = \sqrt{2}(b_3 - b_2) \sin(k_x/2)$ ,
$h_{24} = 
          - \sqrt{2} b_1 \sin (\frac{k_x+k_y}{2})
           + \sqrt{2} b_4 \sin (\frac{k_x-k_y}{2})$
$h_{34} =  2 a_1 \cos (k_y/2) $. 
The existence of such a real representation 
implies that each eigenband state $|E_{\alpha}(\bm{k})\rangle$ ($\alpha=1,\cdots,4$)
can be decomposed as
\begin{subequations}
\begin{eqnarray}
|E_{\alpha}(\bm{k})\rangle = \sum_{j=1}^{4}  p_j(\bmk) ^{1/2} d^{\alpha}_{j}(\bm{k}) |j \bm{k}\rangle \; .
\label{eigen_function1}
\end{eqnarray}
where each component $d^{\alpha}_{j}(\bm{k})$ is a real valued quantity.
In the following we denote 
$E_1(\bm{k})(> E_{2}(\bm{k})> E_{3}(\bm{k})>E_{4}(\bm{k}))$  the conduction band of a $3/4$ filled system.

From this point, the first step to demonstrate the existence of NLs consists 
to show that because of inversion symmetry, at a given TRIM, for each band $E_{\alpha}(\bm{k})$, 
some of the components $d^{\alpha}_{j}(\bm{G}/2)$ necessarily take a value zero. 
Consider any $\bm{G}/2$ TRIM, by construction each basis state $|j \bm{G}/2\rangle$ verifies  $\hat P |j \bm{G}/2\rangle = p_j(\bm{G}/2) |j \bm{G}/2\rangle$
with $p_j(\bm{G}/2)=+1$ $(-1)$ for an even (odd) parity state;  more quantitatively 
$p_1(\bm{G}/2)=(+ + + +)$, $p_2(\bm{G}/2)=(----)$,  $p_3(\bm{G}/2)=(+-+-)$ and $p_4(\bm{G}/2)=(+--+)$ for $\bm{G}/2=(\Gamma,X,Y,M)$ respectively.
Owing to $[P(\bm{G}/2),H(\bm{G}/2)]=0$, each $|E_{\alpha}(\bm{G}/2)\rangle$ is also an eigenstate
of $\hat P$,
\begin{eqnarray}
 \hat P |E_{\alpha}(\bm{G}/2)\rangle = \pi_{\alpha}(\bm{G}/2) |E_{\alpha}(\bm{G}/2)\rangle \; ,
\end{eqnarray}
\end{subequations}
with $\pi_{\alpha}(\bm{G}/2)=\pm 1$.
From this property, it is immediate to deduce that $d_j ^{\alpha}(\bm{G}/2)=0$ for $p_j(\bm{G}/2)=-\pi_{\alpha}(\bm{G}/2)$
whereas  $d_j ^{\alpha}(\bm{G}/2)$  can be finite for $p_j(\bm{G}/2)=\pi_{\alpha}(\bm{G}/2)$.
As a consequence, for each $\alpha$, it is easily checked that if
$\Pi_{\alpha}=\pi_{\alpha}({\bf \Gamma})\pi_{\alpha}({\bf X})\pi_{\alpha}({\bf Y})\pi_{\alpha}({\bf M})=+1$ then for each $j$
there is necessarily an even number of TRIM where $d_j ^{\alpha}(\bm{G}/2)=0$ whereas if $\Pi_{\alpha}=-1$ then for each $j$ there is  
an odd number of TRIM where $d_j ^{\alpha}(\bm{G}/2)=0$. 
\begin{table}

\begin{center}
\begin{minipage}{0.46 \linewidth}
\begin{tabular}{ccccc}
\multicolumn{5}{c}{(a) \ \ \ $\Pi_1=-1$}\\
\hline\noalign{\smallskip}
 & $\Gamma$ & X & Y & M \\
\noalign{\smallskip}\hline\noalign{\smallskip}
$\pi_1(\bm{G}/2)$ & + & - & + & + \\
$d_1^{1}(\bm{G}/2)$ & f & 0 & f & f \\
$d_2^{1}(\bm{G}/2)$ & 0 & f & 0& 0\\ 
$d_3^{1}(\bm{G}/2)$ & f & f & f & 0 \\
$d_4^{1}(\bm{G}/2)$ & f & f & 0 & f \\
\noalign{\smallskip}\hline
\end{tabular}
\end{minipage} \ \ 
\begin{minipage}{0.46 \linewidth}
\begin{tabular}{ccccc}
\multicolumn{5}{c}{(b) \ \ \ $\Pi_{1}=+1$}\\
\hline\noalign{\smallskip}
 & $\Gamma$ & X & Y & M \\
\noalign{\smallskip}\hline\noalign{\smallskip}
$\pi_1(\bm{G}/2)$ & - & - & + & + \\
$d_1^{1}(\bm{G}/2)$ & 0 & 0 & f & f \\
$d_2^{1}(\bm{G}/2)$ & f & f & 0& 0\\ 
$d_3^{1}(\bm{G}/2)$ & 0 & f & f & 0 \\
$d_4^{1}(\bm{G}/2)$ & 0 & f & 0 & f \\
\noalign{\smallskip}\hline
\end{tabular} 
\end{minipage}
\end{center}
\caption{Parities $\pi_1(\bm{G}/2)$ and components $d_j ^{1}(\bm{G}/2)$ of state $|E_{1}(\bm{G}/2) \rangle$. 
We note $d_j ^{1}(\bm{G}/2)=$ f (finite) for $p_j(\bm{G}/2)= \pi_{1}(\bm{G}/2)$ and $d_j ^{1}(\bm{G}/2)=0$ for $p_j(\bm{G}/2)= -\pi_{1}(\bm{G}/2)$. 
\suzu{Parameters of transfer energies are given in the main body.}
Case (a), the transfer energies are chosen such that $\Pi_1=-1$.
In that situation valence and conduction bands cross at a pair of Dirac points $\pm \bmk_0$ and for each j, $d_j ^{1}(\bm{G}/2)=0$ at an odd number of TRIM.
Case (b) corresponds to $\Pi_1=+1$; there is no band crossing between valence and conduction bands and for each j, $d_j ^{1}(\bm{G}/2)=0$ at an even number of TRIM.
}

\label{table_II}
\end{table}
To  illustrate these properties, in Table \ref{table_II} 
we show the WF components $d_j ^{1}(\bm{G}/2)$ of conduction band state $|E_{1}(\bm{G}/2) \rangle$ for two cases.
In case (a), corresponding to uniaxial pressure 
$P_a=6$ kbar \cite{Katayama2006_JPSJ75}, the transfer energies are given by  $a_1, a_2, \cdots, b_4$ = 
 -0.043, -0.096, 0.017, 0.123, 0.149, 0.074 and 0.025 (eV).
In this situation $\Pi_1=-1$, 
and according to our previous work \cite{Piechon2013_JPSJ}, using Fu-Kane topological argument \cite{Fu2007_PRB76}, it implies that
valence and conduction bands cross at a pair  Dirac points $\pm \bmk_0$.
In case (b) $a_1, a_2, \cdots, b_4$ = -0.140, -0.408, -0.002, 0.123, 0.209, 0.151 and 0.025 (eV) corresponding to 
$P_a=45 $ kbar.
The Dirac points have now merged and there is a full gap separating valence and conduction band, such that $\Pi_1=+1$.
\cite{Kobayashi2007_JPSJ76}

We now explain how a local zero $d_j ^{\alpha}(\bm{G}/2)=0$ necessarily implies the existence of a NLs $d_{j} ^{\alpha}(\bmk)=0$.
Let us define $\bmk_{\pm}=\bm{G}/2\pm \bmq$. Owing to time reversal symmetry we have the equality $E_{\alpha}(\bmk_+)=E_{\alpha}(\bmk_-)$.
In addition, because of inversion symmetry  we can also rewrite 
\begin{subequations}
\label{extended_parity}
\begin{equation}
 {H}(\bmk_{\pm}) = H_{\bm{G}/2}^s(\bmq)\pm H_{\bm{G}/2}^a(\bmq),
\label{extended_paritya}
\end{equation}
\begin{equation}
\textrm{with} \ \ \ \
\begin{array}{l}
[P(\bm{G}/2),H_{\bm{G}/2}^s (\bmq)]=0,\\
\lbrace P(\bm{G}/2),H_{\bm{G}/2}^a (\bmq)\rbrace=0.
\end{array}
\label{extended_parityb}
\end{equation}
\end{subequations}
As an example, for $\bmk_{\pm}={\bf Y}\pm \bmq$,
the non vanishing elements of $H_{{\bf Y}}^{s}(\bmq)$ are $h_{11},h_{13},h_{22}=-h_{11},h_{24}$, those of $H_{{\bf Y}}^{a}(\bmq)$ are 
$h_{12},h_{14},h_{23},h_{34}$ and properties eq. (\ref{extended_parityb}) are easyly checked.
Using eq. (\ref{extended_paritya},\ref{extended_parityb}) and equality $E_{\alpha}(\bmk_+)=E_{\alpha}(\bmk_-)$
we deduce the following parity properties for WF components $d_{j} ^{\alpha}(\bmk_{\pm})$:
\begin{subequations}
\label{extended_parity}
\begin{eqnarray}
d_{j} ^{\alpha}(\bmk_+)=d_{j} ^{\alpha}(\bmk_-) \ \ \ \textrm{for } \ \ \  p_j(\bm{G}/2)=\pi_{\alpha}(\bm{G}/2),\\
d_{j} ^{\alpha}(\bmk_+)=-d_{j} ^{\alpha}(\bmk_-) \ \ \ \textrm{for } \ \ \ p_j(\bm{G}/2)=-\pi_{\alpha}(\bm{G}/2).
\label{extended_parityc}
\end{eqnarray}
\end{subequations}
Let us now examine the consequence of the last property eq. (\ref{extended_parityc}) 
in the context of our Hamiltonian that considers only nearest-neigbhor hoppings.

To begin with, we consider the case of a band $E_{\alpha}(\bmk)$ that does not cross any other band and such that $\Pi_{\alpha}=+1$.
In that situation the $d_{j} ^{\alpha}(\bmk)$ can be considered as continuous quantities
 in the entire BZ and therefore property eq. (\ref{extended_parityc}) implies that
each $\bm{G}/2$ TRIM with $p_j(\bm{G}/2)=-\pi_{\alpha}(\bm{G}/2)$ gives rise to a NL $d_{j} ^{\alpha}(\bmk_{\pm})=0$ starting from $\bm{G}/2$.
In addition eq. (\ref{extended_parityc}) also demands that the number of NLs  $d_{j} ^{\alpha}(\bmk_{\pm})=0$ 
that go through a single $\bm{G}/2$ TRIM is necessarily odd. As a consequence a NL cannot be a closed loop inside the BZ
and therefore 
it eventually crosses the zone boundary before reaching back $\bm{G}/2$.
Since $\Pi_{\alpha}=+1$, for a given $j$ there is an even number of $\bm{G}/2$ TRIM  with  $p_j(\bm{G}/2)=-\pi_{\alpha}(\bm{G}/2)$
such that we may expect many NLs originating from distinct TRIM. All these NLs eventually lead to a complicated pattern of sign change 
 for the considered component $d_{j} ^{\alpha}(\bmk)$. This scenario might certainly happend for Hamiltonian with long range hoppings, 
however for our case with only nearest-neighbors hoppings we always obtain
 that a NL $d_{j} ^{\alpha}(\bmk_{\pm})=0$ starts from a given $\bm{G}/2$ TRIM 
and crosses the zone boundary precisely at another $\bm{G'}/2$ TRIM such that $p_j(\bm{G'}/2)=-\pi_{\alpha}(\bm{G'}/2)$.
In other words, when $\Pi_{\alpha}=+1$, NLs connect two partners $\bm{G}/2$ TRIM.
As an example, for the case (b) of Table \ref{table_II} there is
 one NL for each component such that $d_{1} ^{1}(\bmk_{\pm})=0$ connects $\Gamma$ to $X$, 
$d_{2} ^{1}(\bmk_{\pm})=0$ connects $Y$ and $M$, $d_{3} ^{1}(\bmk_{\pm})=0$ connects $\Gamma$
 to $M$, $d_{4} ^{1}(\bmk_{\pm})=0$ connects $X$ and $Y$.

\begin{figure}
  \centering
\includegraphics[width=6cm]{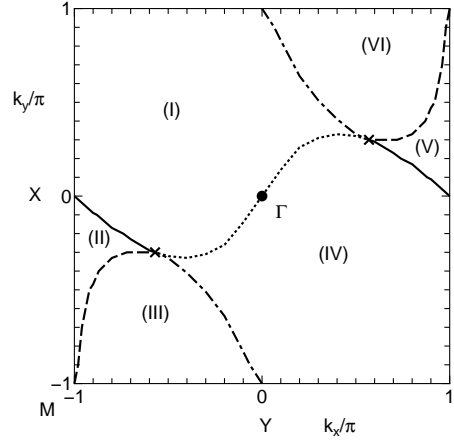}   
  \caption{NLs $d_{j=1,2,3,4} ^1(\bmk_{\pm})=0$ of band $E_1(\bmk)$ in case (a) such that $\Pi_1=-1$
 where transfer energies are the same as Table \ref{table_II}. 
$d_1 ^1(\bmk_{\pm})= 0$ starts from ${\bf X}$ point (solid line), 
$d_2 ^1(\bmk_{\pm})= 0$ starts from ${\bf \Gamma}$ point (dotted line), 
$d_3 ^1(\bmk_{\pm})= 0$ starts from ${\bf M}$ (dashed line) and  
$d_4 ^1(\bmk_{\pm})= 0$ starts from ${\bf Y}$ (dot dashed).
The four NLs $d_{j=1,2,3,4} ^1(\bmk_{\pm})=0$ meet each other at the Dirac points $\pm \bmk_0$ (with $\bm{k}_0 = (0.57,0.3)\pi$) 
and define six regions regions (I), $\cdots$, (VI). 
The NL $d_2 ^1(\bmk_{\pm})= 0$ that connects $Y$ to $M$ is not shown here.
}
\label{NL}
\end{figure}

\begin{figure}
\begin{tabular}{cc}
\includegraphics[height=35mm]{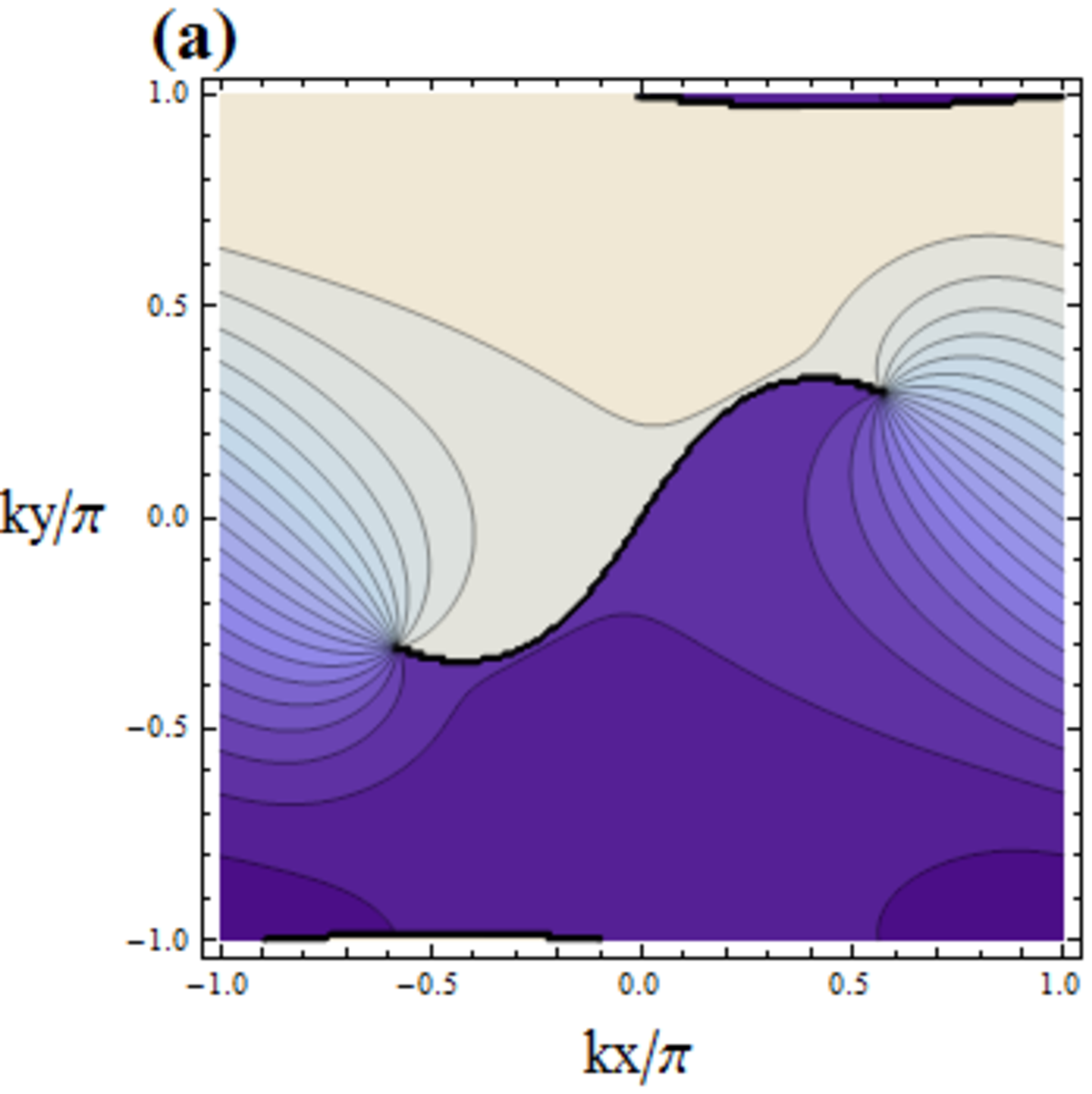}&
\includegraphics[height=35mm]{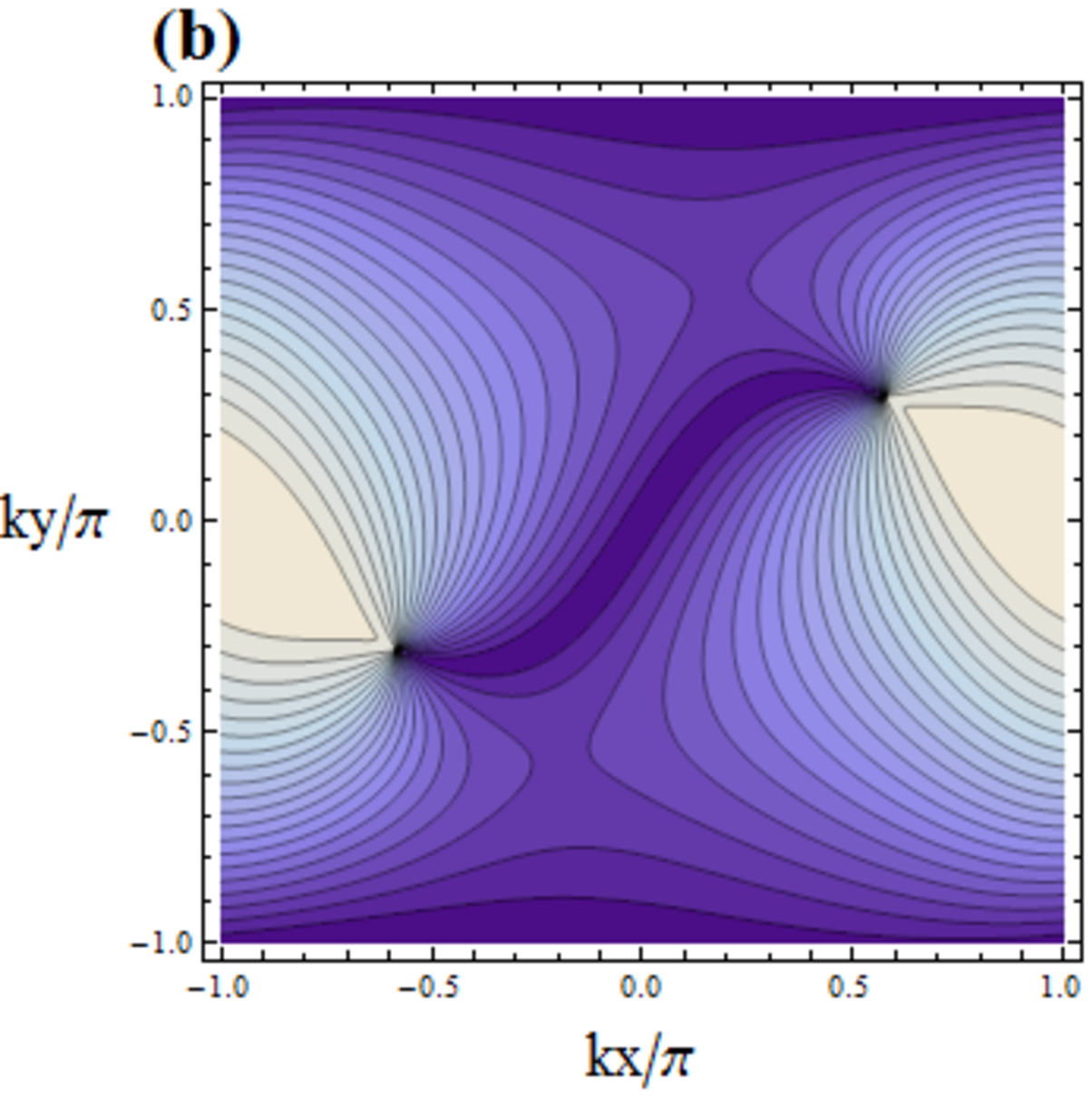}\\
\includegraphics[height=35mm]{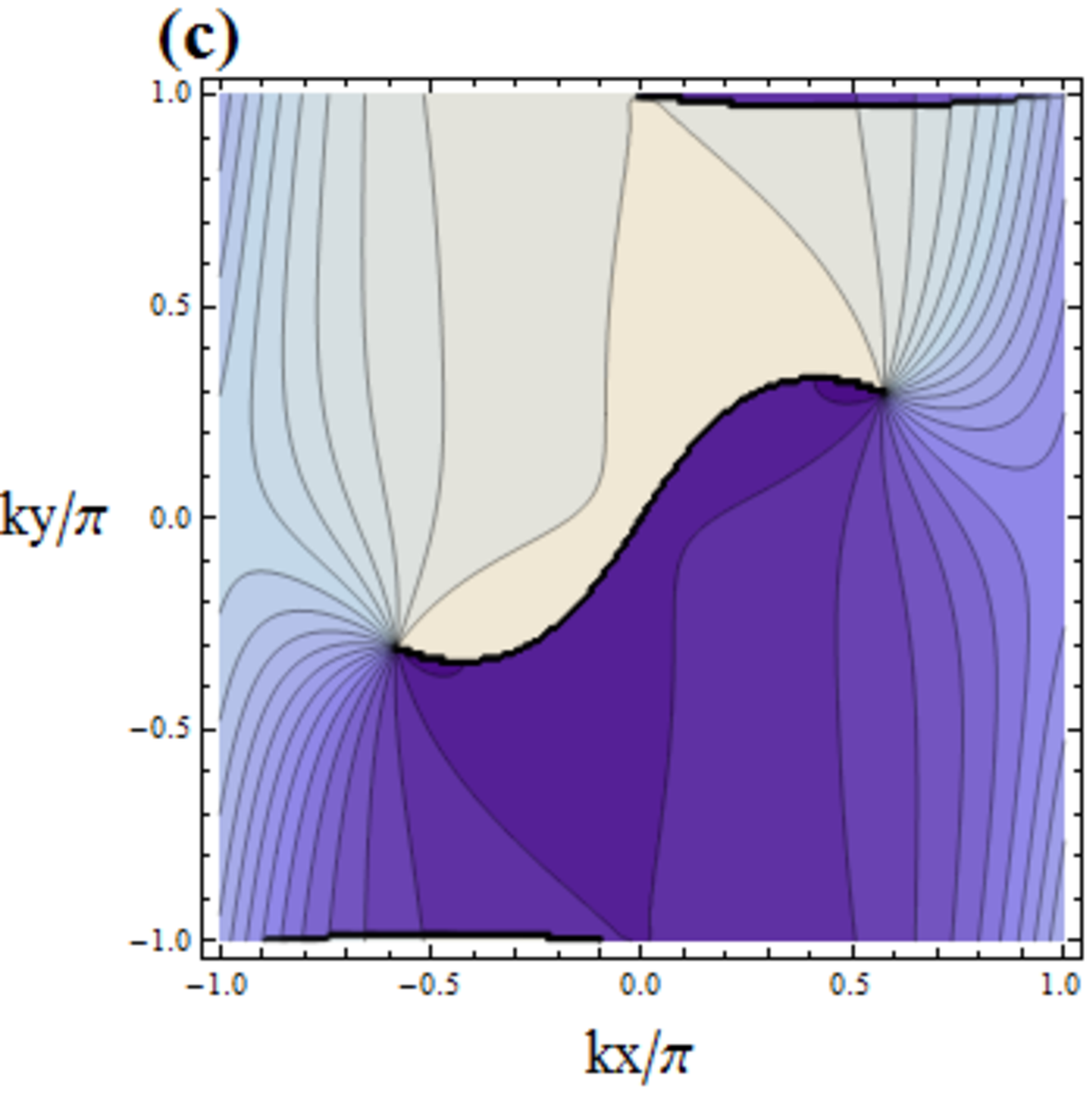}&
\includegraphics[height=35mm]{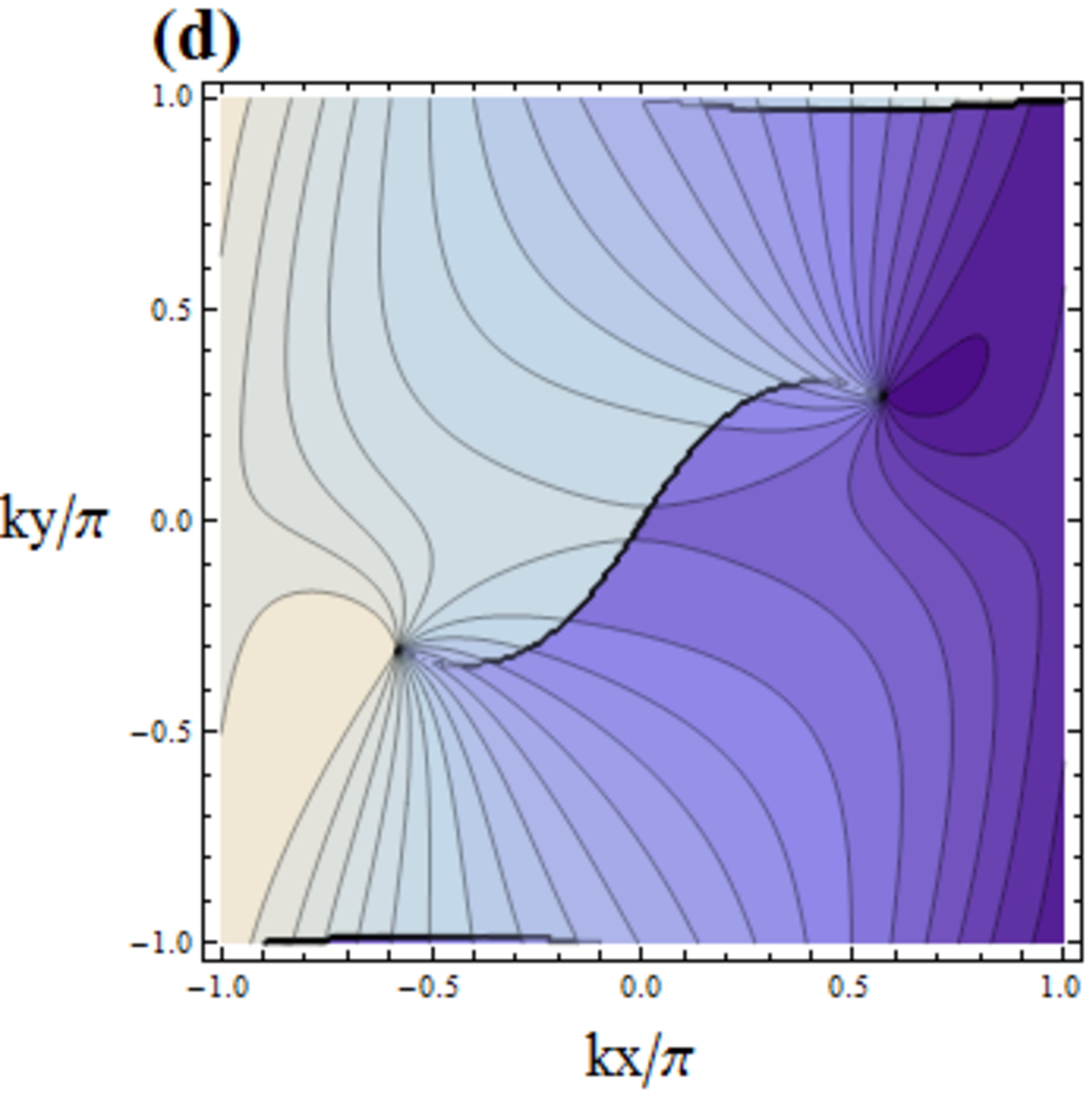}
\end{tabular}
\caption{\label{fig:zerolines}(Color online)
Contour plots for  
$d^1_1(\bm{k})$ (a), 
$d^1_2(\bm{k}) (>0)$ (b), 
$d^1_3(\bm{k})$ (c), and
$d^1_4(\bm{k})$ (d)
 with  
$\pm \bm{k}_0$ $= \pm (0.57,0.3)\pi$.
At the cut  shown by the bold line, components
$d_j(\bm{k})$ ($j$ = 1,3,4) change  sign discontinuously but keep their modulus.
The respective contours take values    
from 0.0756 to -0.0756 (a),
from 0.78 to 0.039 (b),
from 0.0684 to -0.0684 (c),
and
from 0.064 to -0.064 (d).
}
\label{Fig:Fig2}
\end{figure}

We now consider the case of a band 
$E_{\alpha}(\bmk)$ that crosses a neighboring band 
at Dirac points $\pm \bmk_0$; such that $\Pi_{\alpha}=-1$.
In that situation for each $j$ there is an odd number of $\bm{G}/2$ TRIM where $p_j(\bm{G}/2)=-\pi_{\alpha}(\bm{G}/2)$
and therefore there exists necessarily a NL $d_{j} ^{\alpha}(\bmk_{\pm})=0$ that starts from a $\bm{G}/2$ TRIM without a partner $\bm{G'}/2$
 TRIM to connect with. Our numerical calculations show that such a NL $d_{j} ^{\alpha}(\bmk_{\pm})=0$ (without partner) starts from a $\bm{G}/2$ 
and terminates at the Dirac points $\pm \bmk_0$\cite{alternative}. This implies that for each $j$, by going around $\pm \bmk_{0}$ 
it is now possible to go from a region 
$d_{j} ^{\alpha}(\bmk_+)<0$ to a region $d_{j} ^{\alpha}(\bmk_-)>0$ without crossing the NL $d_{j} ^{\alpha}(\bmk_{\pm})=0$.
Such a possibility necessarily implies that $d_{j} ^{\alpha}(\bmk)$ has some discontinuous sign change in the BZ. 
Since all such NLs $d_{j} ^{\alpha}(\bmk_{\pm})=0$ meet each other at $\pm \bmk_0$, we expect the discontinuities of a component $d_{j} ^{\alpha}(\bmk)$ 
to be located on the other NLs $d_{j'\ne j} ^{\alpha}(\bmk_{\pm})=0$.
As an example, we consider the case (a) of Table \ref{table_II}. Reading Table \ref{table_II}(a) we expect that the four NLs
$d_{1,2,3,4} ^1(\bmk_{\pm})= 0$ that start respectively from ${\bf X},{\bf \Gamma},{\bf M}$ and ${\bf Y}$ 
to meet each other at the Dirac points $\pm \bmk_0$ with $\bm{k}_0 = (0.57,0.3)\pi$.
As shown in Figure \ref{NL}, these four NLs  allow  to define six regions (I), $\cdots$, (VI).
In each region, each component $d_{j} ^1(\bmk)$ is continuous and has a given sign. 
Moreover eq. (\ref{extended_parityc}) demands that $d_{1} ^1(\bmk)$ changes sign continuously 
in going from region (I) to (II) or from region (IV) to (V), 
 and similarily for \suzum{$d_{2} ^1(\bmk)$ in going from region (I) to (IV),} 
 $d_{3} ^1(\bmk)$ in going from region (II) to (III) or from region (V) to (VI) and
 $d_{4} ^1(\bmk)$ in going from region (III) to (IV) or from region (VI) to (I).
 For component $d_{2} ^1(\bmk)$ from Table \ref{table_II}(a) we also deduce the existence of another NL $d_2 ^1(\bmk_{\pm})= 0$
that connects $Y$ to $M$. This latter NL implies that $d_{2} ^1(\bmk)$ does not change sign in going from (III) to (IV) or from (I) to (VI).
All these properties appear however insufficient to determine uniquely the sign of the components $d_{1,2,3,4} ^1(\bmk)$ in the six regions.
To this end we use explicit numerical calculations. 
Since the global sign of the WF components is not determined, 
hereafter we choose to set  $d^1_2(\bm{k}) \ge 0$.
In Table \ref{table_III}, within this convention, we present the sign of $d^1_j(\bm{k})$ ($j=1,2,3,4$)
 in the six regions shown in Fig. \ref{NL}. 
 We observe that the sign changes of components $d^1_j(\bm{k})$ ($j=1,3,4$) respect the properties implied by their respective NLs.
\begin{table}
\caption{Sign of $d^1_{1,2,3,4}(\bm{k})$ for the six regions shown in Fig.~\ref{NL}
(Case (a) for band $E_1(\bmk)$ such that $\Pi_1=-1$). }
\begin{center}
\begin{tabular}{cccccccc}
\hline\noalign{\smallskip}
   & (I) & (II)& (III)& (IV) &(V) & (VI) \\
\noalign{\smallskip}\hline\noalign{\smallskip}
$d^1_1(\bm{k})$ & + & - & - & - & + & +  \\
$d^1_2(\bm{k})$ & + & + & + & + & + & +  \\
$d^1_3(\bm{k})$ & + & + & - & - & - & +  \\
$d^1_4(\bm{k})$ & + & + & + & - & - & - \\
\noalign{\smallskip}\hline
\end{tabular}
\end{center}
\label{table_III}
\end{table}
In order to understand Table \ref{table_III} more explicitly, 
the contours plots of $d^1_{1,2,3,4}(\bm{k})$  are shown in Fig.~\ref{Fig:Fig2}.
Since $d^1_2(\bm{k})$ is chosen to be  positive,
for the other components $d^1_j(\bm{k})$ ($j=$ 1, 3 and 4 ) there appears a cut
along the NL $d^1_2(\bm{k}_{\pm})=0$.
At the cut these components 
 $d^1_j(\bm{k})$ ($j=$ 1, 3 and 4 ) exhibit a discontinuous sign change but keep the same modulus.
The dark regions (bright region) in (a), (c) and (d) correspond 
to $d^1_j(\bm{k}) < 0 ( > 0 )$.

Using Fig.~\ref{Fig:Fig2},  we examine the Berry phase,  $\gamma_C$, defined  by
\cite{Berry1984} 
 \begin{eqnarray}
  \gamma^1_C &= & 
 {\rm Im}  \int_{\rm C} 
  \langle E_1(\bm{k})| {\nabla}_{\bm{k}} |E_1(\bm{k})  \rangle
\cdot {\rm d} \bm{k} 
\label{eq:Berry_C}
\end{eqnarray}
 where $C$ denotes a closed loop taken anti-clock wise. From eq. \ref{eigen_function1}, 
 we note that the phases contained in $p_j(\bmk)^{1/2}$ cannot contribute to any finite Berry phase.
 In fact in our case all the possible finite Berry phase contributions 
 come from the discontinuous sign change that occurs for the components $d^1_j(\bm{k})$ ($j$ =1, 3 and 4) 
 at the cut located on the NL $d^1_2(\bm{k}_{\pm})=0$ that connects the Dirac points via 
 the $\Gamma$ point.
 Indeed, even if  $d^1_{j}(\bm{k})$ ($j$ =1, 3 and 4)  are real quantities, each discontinuous sign change  corresponds 
 to a phase jump of $\theta^1_j(\bm{k}_{C^+})- \theta^1_j(\bm{k}_{C^-}) =\pm \pi$ across the cut 
(i.e.  a singular gradient contribution of the phase)\cite{Montambaux2011_PRB} 
 such that we can write
\begin{equation}
 \begin{array}{ll}
 \gamma^1_C&= {\rm Im}  
\sum_{j=1}^4 \int_{\rm C}
   d^1_j(\bm{k}) {\nabla}_{\bm{k}} d^1_j(\bm{k})  \
\cdot {\rm d} \bm{k} \\ 
 &=\sum_{j=1}^4 
  |d^1_j(\bm{k}_C)|^2  (\theta^1_j(\bm{k}_{C^+})- \theta^1_j(\bm{k}_{C^-}) ),
\end{array}
\label{eq:Berry_C1}
\end{equation}
where $\bm{k}_{C^-}$ ($\bm{k}_{C^+}$ ) denotes a point located just below (above) the cut constitued by the NL  $d^2_j(\bm{k}_{\pm})=0$. 
In the present case, we take  
 $\theta^1_j(\bm{k})=0$ for $d^1_j(\bm{k})>0$ and 
  $\theta^1_j(\bm{k})=  \pi$ for $d^1_j(\bm{k})<0$ 
 for each region of Fig.~\ref{Fig:Fig2}.
  By noting that $\sum_j ({d^1_j}(\bm{k}))^2 = 1$, and that $d^1_2(\bm{k})=0$ on the cut,
 we obtain $\gamma^1_C=\pi$
when $C$ encloses one of the Dirac points $\pm \bm{k}_0$
and $\gamma^1_C=0$ when $C$ encloses either two or zero Dirac points $\pm \bm{k}_0$.

In summary, in the context of a tight-binding model of  organic conductor $\alpha$-(BEDT-TTF)$_2$I$_3$,
we have shown that inversion and time reversal symmetries implies the existence of NLs for WFs components written 
in the inversion Bloch state basis. There exists two kinds of NLs. On the one hand there are NLs that connect two partners TRIM, 
 on the other hand when there exists a pair of Dirac points, for each component there appears a NL that connects the pair of Dirac points via a single TRIM. 
 The NLs of this second type, are also the location of discontinuous sign change of WFs components (i.e. phase jump of $\pm \pi$) 
 which is at the origin of the Berry phase accumulated when encircling a Dirac point. 
 Interesting perspectives would consist to examine how to generalize these results to other type of crystal symmetries. 

\acknowledgements
We thank A. Kobayashi, T. Morinari, and T. Tohyama for useful discussions.
One of us (Y.S.)  thanks T. Kariyado for useful comments on the Berry phase.
This work was supported 
 by a Grant-in-Aid for Scientific Research (A)
(No. 24244053) and 
(C)  (No. 23540403)
 from the Ministry of Education, Culture, Sports, Science, and Technology, Japan,




\begin{thebibliography}{}

\bibitem{Mori1984_CL} 
T. Mori, A. Kobayashi, T. Sasaki, H. Kobayashi, G. Saito, 
 and H. Inokuchi: Chem. Lett. (1984) 957. 

\bibitem{Katayama2006_JPSJ75} 
S. Katayama, A. Kobayashi, and Y. Suzumura: J. Phys. Soc. Jpn. \textbf{75} (2006) 054705.

\bibitem{Tajima2009_STAM10} 
N. Tajima and K. Kajita: Sci. Technol. Adv. Mater. {\bf 10} (2009) 024308. 

\bibitem{Kobayashi2009_STAM10}
A. Kobayashi, S. Katayama, and Y. Suzumura: 
Sci. Technol. Adv. Mater. {\bf 10} (2009) 024309.


\bibitem{Fu2007_PRB76} 
L. Fu and C. L. Kane:
Phys. Rev. B {\bf 76} (2007) 045302.

\bibitem{Piechon2013_JPSJ} 
F. Pi\'echon and Y. Suzumura: J. Phys. Soc. Jpn. \textbf{82} (2013) 033703.


\bibitem{Mori2013_JPSJ} 
T. Mori: J. Phys. Soc. Jpn. \textbf{82} (2013) 034712.

\bibitem{Ando2005_JPSJ74} 
For example, see  review by T. Ando: J. Phys. Soc. Jpn {\bf 74} (2005) 777.


\bibitem{Mori2010_JPSJ} 
T. Mori: J. Phys. Soc. Jpn. \textbf{79} (2010) 014701.


\bibitem{Suzumura2011_JPSJ_Berry} 
Y. Suzumura and A. Kobayashi: J. Phys. Soc. Jpn. \textbf{80} (2011) 104701.

\bibitem{Suzumura2013_JPSJ} 
Y. Suzumura, T. Morinari and F. Pi\'echon: J. Phys. Soc. Jpn. \textbf{82} (2013) 023708.


\bibitem{Katayama2009_EPJB57}
S. Katayama, A. Kobayashi, and Y. Suzumura: 
Eur. Phys. J. B. {\bf 67} (2009) 139.


\bibitem{Kobayashi2013_JPSJ} 
A. Kobayashi and Y. Suzumura: J. Phys. Soc. Jpn. \textbf{82} (2013) 054715. 

\bibitem{Takano2010_JPSJ79}
Y. Takano, K. Hiraki, Y. Takada,  H. M. Yamamoto, and T. Takahashi:
   J. Phys. Soc. Jpn. {\bf 79} (2010) 104604.

\bibitem{Hirata2012}
M. Hirata: Ph.D. thesis, The University of Tokyo, (2012).


\bibitem{Kobayashi2007_JPSJ76}
 A. Kobayashi, S. Katayama,  Y. Suzumura, and H. Fukuyama: J. Phys. Soc. Jpn. \textbf{76} (2007) 034711.


\fred{\bibitem{alternative} An alternative view that is confirmed by our numerical calculations is the following:
 for two bands $E_{\alpha}(\bmk),E_{\alpha'}(\bmk)$ that cross at Dirac points $\pm \bmk_0$, for each $j$ there is a NL  
$d_j^{\alpha}(\bmk_{\pm})=0$ ($d_j^{\alpha'}(\bmk_{\pm})=0$) that start from a ${\bm G}/2$ (${\bm G}'/2$) TRIM and terminates at $\pm \bmk_0$. 
Since ${\bm G} \ne {\bm G}'$, it appears that the reunion of these two NLs now constitutes a NL $d_j^{\alpha \alpha'}(\bmk_{\pm})=0$ that connects two partners TRIM.}

\bibitem{Berry1984}
M. V. Berry: 
Proc. R. Soc. Lond. A \textbf{392} (1984) 45.


\bibitem{Montambaux2011_PRB} 
D. Deplace, D. Ullmo and  G. Montambaux:
Phys. Rev B  {\bf 84} (2011) 195452.





\end{thebibliography}
\end{document}